\documentclass[aps,prl,reprint,twocolumn,superscriptaddress,floatfix]{revtex4-1}

\usepackage{amsmath}
\usepackage{graphicx}
\usepackage{dcolumn}
\usepackage{mathrsfs}
\usepackage{multirow}
\usepackage{mciteplus}
\usepackage{color}
\usepackage{ulem}
\usepackage{epstopdf}

\newcommand\ket[1]{\vert #1 \rangle}

\newcommand\braket[2]{\langle #1 \vert #2 \rangle}

\begin{document}

%\preprint{}

\title{Laser pulses for coherent xuv Raman excitation}
%\\using quantum
 % optimal control  
 % and multichannel electronic structure theory}


\author{Loren Greenman}
\affiliation{Department of Chemistry and Kenneth S. Pitzer Center for
  Theoretical Chemistry, University of California, Berkeley, CA 94720,
  USA} 
\affiliation{Chemical Sciences, Lawrence Berkeley National Laboratory,
  Berkeley, CA 94720, USA} 
\author{Christiane P. Koch}
\affiliation{Theoretische Physik, Universit\"at Kassel,
  Heinrich-Plett-Str. 40, D-34132 Kassel, Germany} 
\author{K. Birgitta Whaley}
\email{whaley@berkeley.edu}
\affiliation{Department of Chemistry and Kenneth S. Pitzer Center for
  Theoretical Chemistry, University of California, Berkeley, CA 94720,
  USA} 
\affiliation{Chemical Sciences, Lawrence Berkeley National Laboratory,
  Berkeley, CA 94720, USA} 

\date{\today}


\begin{abstract}
We combine 
%the time-dependent configuration interaction singles (TDCIS) method
multi-channel electronic structure theory 
with quantum optimal control to
derive Raman pulse sequences that coherently populate a 
valence excited state. For a neon atom, 
Raman target populations of up to 13\% 
are obtained. 
Superpositions of the ground and valence Raman states
with a controllable relative phase are found to be reachable with up to
4.5\% population and phase control facilitated by the pump pulse
carrier envelope phase. Our results
open a route to creating core-hole excitations in molecules and
aggregates that locally address specific atoms and represent the
first step towards realization of multidimensional spectroscopy in the
xuv and x-ray regimes. 
\end{abstract}

\maketitle

Multidimensional spectroscopy in the
infrared~\cite{larsen2001vibrations} and
uv-vis~\cite{mukamel2000mdfemto,jonas2003twodfemto} spectral regions
has proven to be a powerful tool for revealing quantum coherent dynamics in biological
systems~\cite{engel2007fmo} and quantum
devices~\cite{li2006semiconductor,erementchouk2007semiconductor,yang2008semiconductor}.
% allowing, e.g., for analysis of excitation
% pathways~\cite{mukamel2000mdfemto,jonas2003twodfemto}. 
%%% chr: correct? 
%%% lg: what is in question here?
%BW will have to check refs 
Extending the techniques of multidimensional spectroscopy to the xuv
and x-ray regimes could open the door to studying 
energy transfer between different atomic sites in
molecules~\cite{tanaka2002coherent,mukamel2009coherent}. 
It would provide a local probe of valence excitations, which would
be invaluable  for studies of energy transfer processes in 
biological systems and quantum devices.
However, 
this presents novel challenges, since the large energy of the xuv and x-ray pulses 
can result in a high probability of ionization, while selective
excitation of a specific intermediate state may be hampered by the
presence of a multitude of other states closely lying by. 

On an abstract level, 
these difficulties reflect the problem of controllability when a
continuum of states is involved~\cite{ShapiroBook2}. 
Controllability addresses the question whether a quantum control
target is reachable, given the properties 
of the Hamiltonian~\cite{DAlessandroBook}. 
For a structureless continuum, no control is expected, whereas
resonances in the continuum are predicted to facilitate
control~\cite{ZemanPRL04}, the extent of which depends on the
resonance lifetime compared to the duration of the pulses. 
Standard controllability analysis \cite{DAlessandroBook} 
cannot account for finite lifetimes and is thus not applicable.
This leaves numerical optimization as the method of choice to investigate
how much control is attainable for a process that necessarily proceeds
via a continuum of states.

Here we combine quantum optimal control theory  with 
%a multichannel electronic structure method in order 
the time-dependent configuration interaction singles (TDCIS) 
description of electronic structure to assess the
feasibility of coherent xuv Raman excitation,
as a first step towards multidimensional spectroscopy
in the x-ray regime~\cite{tanaka2002coherent}.
Achieving such coherent Raman excitation with a high yield is
extremely challenging, due to the presence of the ionization continuum
that is accessible by both pump and Stokes pulses. Furthermore, in
order to probe electronic dynamics of valence excitations in
molecules, the capability to prepare coherent superpositions
of valence states is imperative, in addition to population transfer
into the excited states. The primary control objectives are thus to
both drive population to a specific atomic valence excitation and 
to achieve a coherent superposition of ground state and valence
excitation with controllable relative phase.  
We demonstrate here that optimal control theory allows us to predict
experimentally feasible pulse forms  
that populate the desired state up to 13\% and achieve
superpositions of ground and excited state 
with arbitary relative phase and up to 4.5\% excited state population. 
The optimal control approach not only allows for generating the 
pulse forms but also reveals insight into the physical
mechanism of the optimized excitation process. This, in turn, allows
development of simpler pulse sequences with similar performance but
better compatibility  with experimental
constraints. 
Our work is the first to tackle the problem of the presence of the
electronic continuum by optimal control, adding an important capability
to the growing attempts now underway to tailor multi-electron
dynamics~\cite{KlamrothJCP06,MundtNJP09,CastroPRL12,li2014population}.

\begin{figure}[tb]
\includegraphics[height=2in,width=2in]{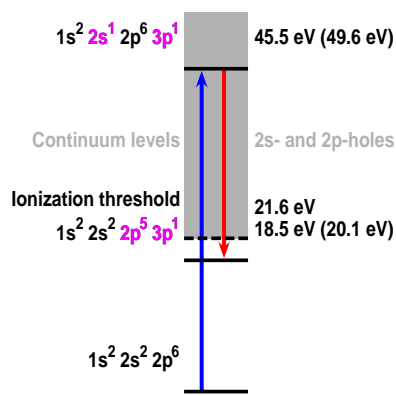}
%\includegraphics[height=2in]{NewerNeRaman}
%\vspace{-2em}
\caption{(Color online) Targeted coherent xuv Raman process in neon.  The
  experimental energies are shown together with TDCIS values in
  parentheses.\label{NeRaman}} 
%  \vspace{-2em}
\end{figure}
We consider as example the neon atom, employing the levels shown in
Fig.~\ref{NeRaman}. They are accessible in table-top experiments
%using a titanium-doped sapphire 800$\,$nm laser that
generating intense high harmonics~\cite{AllisonOL11} 
% in the 5-80$\,$eV range
or using a free-electron laser operating in the xuv regime.
%% Add citation to FLASH or FERMI? 
A pump pulse of 45.5$\,$eV couples the ground state, with
configuration $1s^22s^22p^6$, to the core-excited resonance 
($1s^22s^12p^63p^1$), driving a $2s-3p$ hole-particle
excitation. A Stokes pulse of 27.0$\,$eV induces the filling of the
$2s$ hole with a $2p$ valence electron, creating the $2p-3p$ excitation
($1s^22s^22p^53p^1$) which is the target valence excited sate. 

In order to describe the manifold of excited states and capture the
ionizing electron density, we calculate the quantum dynamics using the
time-dependent configuration interaction singles (TDCIS) method on a
numerical grid with a complex absorbing potential
(CAP)~\cite{greenman2010implementation,xcid}. The TDCIS method was
developed  to capture channel coupling in high harmonic
generation~\cite{greenman2010implementation,Pabst_PRA_12}, and it has
been used to  describe multichannel dynamics in a number of ultrafast
processes~\cite{pabst2011decoherence,wirth2011synthesized,pabst2012krypton,sytcheva2012neon}. 
%add Am J Phys paper and new Stefan paper?
%BW ok to add more if we have space! 
%chr: references don't count towards the page limit
The wavepacket is described by a single-determinant Hartree-Fock
ground state $\ket{\Phi_0}$, single-particle excitations from occupied
orbital $i$ to unoccupied orbital $a$ $\ket{\Phi_i^a}$, and
time-dependent coefficients $\alpha$, 
\begin{equation}
\label{ansatz}
\ket{\Psi (t)}=\alpha_0(t)\ket{\Phi_0}+\sum _{i,a}\alpha _i^a(t)\ket{\Phi _i^a}.
\end{equation}
The dynamical equations for the coefficients are obtained by inserting 
Eq.~(\ref{ansatz}) into the Schr\"odinger
equation~\cite{greenman2010implementation}.  
A complex absorbing potential (CAP) is added to the Hamiltonian in
order to capture 
ionization~\cite{goldberg1978cap,santra2002cap,greenman2010implementation}.

We may simplify the electronic structure calculation by employing a
Hartree-Fock-Slater one-electron potential as a  
%BW  does this mean that you change after one or more iterations? I
%thought the time independent (matter) part of the Hamiltonian was
%constant
% LG 0820 I don't understand quite what you're asking
%         the orbitals are time-independent, the coefficients on them
%         are time-dependent 
%         the orbitals may be taken from Hartree-Fock (full TDCIS)
%                                        Hartree-Fock-Slater (TDCIS-HFS)
%                                        a limited set of
%                                        Hartree-Fock-Slater
%                                        (TDCIS-HFS-1P)  
%         for the HFS-based methods, the Coulomb repulsion is not
%         included in the time propagation 
%            this means it is assumed that the HFS potential
%            approximates the Coulomb potential 
%BW Ok I get it now, this is  different set of calculations with the
%approximate potential 
starting point (TDCIS-HFS).
Dynamical calculations with TDCIS-HFS are found to yield final state
populations agreeing with full TDCIS results to within a factor of
three. 
A further gain in efficiency is possible by using a simplified
configuration space including only ionization levels reachable by
one-photon absorption within a bandwidth of a few eV (TDCIS-HFS-1P). 
%%% chr: please add which calculations use which method, i.e., for
%%% pulse durations longer than XX, 
%%% the propagation is at the TCDIS-HFS-1P level and for pulse
%%% durations longer than YY at the TDCIS-HFS level rather than full TDCIS
%%% lg: This is not really the case.  Of the figures in the paper, only Fig. 4 
%%% shows TDCIS-HFS-1P results.  For all of the other figures, optimizations
%%% were performed with TDCIS-HFS-1P, but propagations were performed
%%% with full TDCIS.  Fig. 5 used to have a TDCIS-HFS result, but now 
%%% has only a full TDCIS result. So I add...
%%% chr: OK!
Pulse sequence optimizations were performed using the TDCIS-HFS-1P method,
and propagations were performed with the optimal pulses at the full TDCIS level.
All calculations employed 1000 grid points in 63.6$\,$\AA , with a CAP
radius of 42.4$\,$\AA~and CAP strength of $10^{-4}$, with angular
momentum functions restricted to $L \le 3$. 
 
Krotov's optimal control
method~\cite{konnov1999krotov,bartana2001krotov,JoseKochPRA08,reich2010monotonically}
is utilized to find pulses which suppress ionization.  It  minimizes
the cost function $J$ for the desired excitation, 
\begin{equation}
\label{JTSS}
J_{t_f}=-\vert\braket{\Phi_D}{\Psi(t_f)}\vert^2, 
\end{equation}
where $\ket{\Phi _D}$ represents the 
target state (the 2p-3p excitation) and $\Psi(t_f)$ the time evolved
state in presence of the external field $\mathcal{E}(t)$. 
The Krotov pulse update formula is given by~\cite{reich2010monotonically}
\begin{equation}
\label{update}
\mathcal{E}(t)^{(k+1)}=\mathcal{E}(t)^{(k)}-\frac{\lambda}{2S(t)}\mathrm{Im}\left(\mathbf{\chi}(t)^T\mathbf{z}\mathbf{\alpha}(t)^{(k+1)}\right),
\end{equation}
where $\mathcal{E}_t^{(k)}$ is the time-dependent field amplitude at
the $k$-th iteration, $S(t)$ an arbitrary shape function which  
ensures that the pulse goes to zero at the ends of the time
propagation, $\mathbf{z}$ the transition dipole matrix in the basis
of TDCIS states, $\mathbf{\alpha}$ a vector of the time-dependent
coefficients of Eq.~\eqref{ansatz}, and $\mathbf{\chi}$ 
the set of corresponding time-dependent co-vectors~\cite{konnov1999krotov}.
The choice of cost function at the final time determines the ``initial
condition''~\cite{konnov1999krotov},
\begin{equation}
\label{conditionchiT}
\mathbf{\chi}(t_f)=2 \mathbf{\alpha}_D \left(
  \mathbf{\alpha}_D^T\mathbf{\alpha}(t_f)\right)^*. 
%\textcolor{red}{FIX ME}
\end{equation}
The covectors $\mathbf{\chi}(t)$ are  propagated backwards in time
according to the equations of motion, 
\begin{equation}
\label{conditionchid}
\dot{\mathbf{\chi}}=i\left(\mathbf{H}^T-\mathcal{E}(t)^{(k)}\mathbf{z}^T\right)
\mathbf{\chi}(t)\,, 
\end{equation}
with $\mathbf{H}$ the time-independent part of the
Hamiltonian~\footnote{The presence of matrix transposes instead of the 
  usual Hermitian conjugates is due to the use of a CAP.}. 
 
\begin{figure}[tb]
\centering
\includegraphics[width=0.99\linewidth]{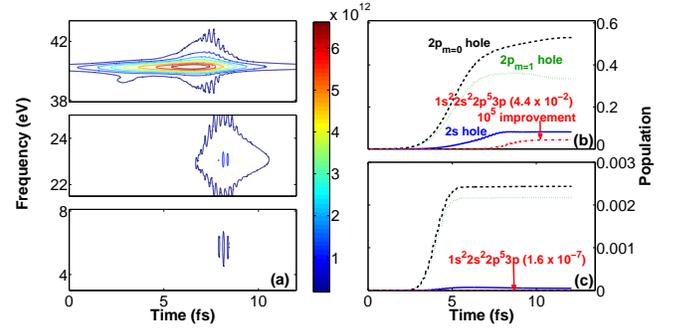}
\caption{(Color online) 
  (a)  Filtered Wigner distribution of the optimized pulse sequence,
  showing the Stokes pulse starting at the end of the pump pulse and
  addition of low frequency (eV) components.  Color bar units are
  W/cm$^2$. (b)  Populations of 2p, 2s hole states and  the
  target $1s^22s^22p^53p$ state that are achieved with the optimized
  pulse. 
  (c) Populations for two simultaneous 2.0$\,$fs TL pulses 
  with central frequencies of 49.6 eV and 29.5 eV.
  The population of the target state (dashed-dotted red line) reaches 
  $4.4 \times 10^{-2}$ for the  optimized pulse, compared to 
  $\sim 1.6 \times 10^{-7}$ for the naive sequence. 
  The 2p holes ($m=0$; black dashed line, $m=1$; green dotted line)
  correspond to $1s^22s^22p^5nl$ configurations,
%BW I don't think we need to say what is excited and in any case I think it is surely 'can excited nl electron' not an excited 2 p orbital?
% with an excited 2p-orbital. 
% LG 0822 are these comments that need to be addressed, I'm confused.  Or are they leftover old comments?
  the 2s hole (blue solid line) to $1s^22s^12p^6nl$ configurations.
\label{naivepulse}}
\end{figure}
Our starting point is a naive pulse sequence for the Raman process
obtained by assuming simultaneous, transform-limited (TL) 
pump (p) and Stokes (S) pulses, 
$\mathcal{E}_{p/S}(t)=\mathcal{E}_{0,p/S}\sin(\omega_{p/S} (t -t_{0,p/S})+ \phi_{p/S})
\exp[-4\ln 2(t-t_{0,p/S})^2/\sigma_{p/S}^2]$,
with
parameters corresponding to the experimental setup of
Ref.~\cite{AllisonOL11}: a full-width at half maximum (FWHM) duration
of 2$\,$fs and a peak intensity of 
$3.5\times 10^{12}\,$W/cm$^2$. This sequence 
does not populate the Raman state significantly,
cf. Fig.~\ref{naivepulse}(c). 
The overall depopulation of the ground state is small, of order
$10^{-3}$. This population is divided between 2p holes
% ($1s^22s^22p^5(nl)$ configurations) 
and the 2s hole. %($1s^22s^12p^6(nl)$ configurations).
% Some excited configurations may contain orbitals that are rotated into
% the complex plane by the CAP, representing ionized electron density. 
The target state is one particular configuration of the 2p hole which
is populated only to a few hundredths
of a percent of the total hole population. 

Starting with the naive pulse sequence, % LG 0821 in Fig.~\ref{naivepulse}(a), 
an optimized sequence is obtained using Krotov's method.
The resulting pulses, shown in Fig.~\ref{naivepulse}(a), 
achieve the Raman excitation with a population of $4.4 \times
10^{-2}$, five orders of magnitude better than the starting pulse. 
The optimized pulse is increased in amplitude by a factor of about 16
and therefore ionizes more of the electron density. It 
nevertheless yields an improvement of three orders of magnitude in the
percentage contribution of the target state to the total hole
probability. The peak intensity of the optimal sequence is about 
$7.9\times 10^{14}\,$W/cm$^2$. 
Analysis of the dynamics under the optimized pulse reveals
the added low frequency contribution (Fig.~\ref{naivepulse}(a)) to be
irrelevant, whereas the relative timing of the pump and Stokes pulse
components are key features. 
In particular, the relative timing of the Stokes pulse, which now
enters only at the end of the pump pulse, is critical. 
As developed further below, this suggests a sequential mechanism that
may be used to further optimize the pulse sequence within experimental
constraints.  

\begin{figure}
\begin{tabular}{cc}
\includegraphics[width=0.99\linewidth]{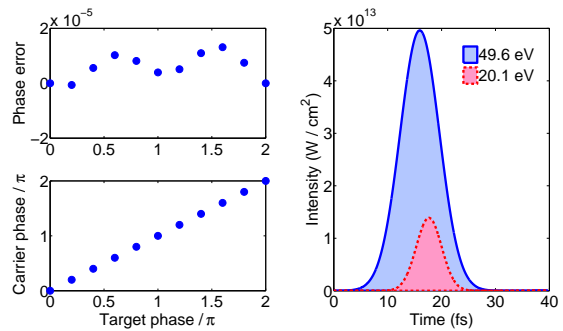}
\end{tabular}
\caption{(Color online) Optimization of coherent superpositions:
  (a) phase error, (b) optimized pump pulse carrier envelope phase
  $\phi_p$ shown as a function of the target phase $\phi_t$, 
  (c) optimized pulse envelope, showing a clear distinction between pump
  (earlier) and Stokes (later) component.   
  \label{phases}}
\end{figure}

% lg 0819  To clarify how the optimization is done, 
%  1.  pick a target phase phi_t
%  2.  run an optimization (all optimizations start from the same pulse)
%  3.  find the result, which
%      a.  has a similar pulse envelope for each optimal pulse, as shown in the figure
%      b.  differs in the carrier envelope phase of the pump pulse
%      c.  the populations of each of the states looks the same for each different target phase
%      d.  the phase between the ground and raman states is the target phase plus a very small error
%
Coherent superpositions of the Raman state with the
ground state are obtained by optimizing with 
$0.99e^{i\phi_t}\ket{1s^22s^22p^53p}+
0.16\ket{1s^22s^22p^6}$ as target state in the cost
function~\eqref{JTSS}, where $\phi_t$ 
is the desired relative phase. 
This target state was empirically determined to drive population to the
Raman excited state while maintaining the target phase with the ground state.
In this type of optimization, 
both amplitude and phase of the pulse sequence are
varied.  This procedure enables identification of  pulses for
arbitrary, prespecified values of the target phase $\phi_t$. 
Figure~\ref{phases}(a) shows the phase error in the target state
as a function of $\phi_t$.
The optimized pulses have a combined peak intensity of about
$10^{14}\,$W/cm$^2$ and  excite a Raman population of 4.5\%
while depopulating the ground state by about 53\%. 
Independent of the value of $\phi_t$, all optimizations are found to
converge on the same pulse envelopes that strongly 
resemble a time-separated pump and Stokes pair, with the Stokes pulse
starting just after the peak of the pump pulse,
cf. Fig.~\ref{phases}(c).  The optimized pulses differ primarily in
the  value of the carrier envelope phase (CEP) of the pump component,
$\phi_p$, calculated using a Hilbert transform of the
pulse~\cite{duoandikoetxea2001fourier}. 
It  is found to correlate closely with the target phase $\phi_t$,
cf. Fig.~\ref{phases}(b).    
Since the propagation of the excited state after the pulse imparts a
phase on the excited state 
that depends on an arbitrary origin of time $t_{origin}$, $e^{-i\omega (t-t_{origin})}$,  
we picked $t_{origin}$ so that a phase of 0.0 was obtained by the pulse with a CEP of 0.0.
%%% lg: you excite a superposition |g> + e^i\phi |x> that propagates as
%%%  |g> + e^i\phi e^-iw(t-t_0) |x>.  The choice of t_0 is arbitrary, since 
%% % it is the origin of time, and you may as well start at 0, -50, 50.
%%% So I picked t_0 so that the CEPs and phases \phi lined up with \phi_t. 
%%% Changing above to clarify.  But maybe we should eliminate for space?
%%% chr: It's clear now but I agree the preceding sentence can be eliminated
The correlation between target phase and pump CEP is rationalized by
the pump imprinting its CEP onto the intermediate state, which yields
the desired relative phase between intermediate and Raman state. 
These results show that a coherent
superposition can be excited with any  desired relative phase $\phi_t$
merely by changing the pump pulse CEP $\phi_p$,
once the optimized pulse envelope has been determined. 

Our optimal control calculations suggest a sequential mechanism
whereby first the intermediate state is 
populated and then the second pulse component acts to transfer
population and phase information to the desired state. Exploiting this
intuition, the pulse sequences can be engineered further by optimizing
each step individually, in order to (i) explore large areas of pulse
parameter space for maximum performance or (ii) obtain simple pulses
with near-optimal performance that are also consistent with experimental
constraints. 
Optimizing a Gaussian-shaped Stokes pulse starting from the intermediate state, we
find complete population transfer to the valence state for a pulse
duration of 0.5$\,$fs and a peak intensity of $2.4 \times
10^{15}\,$W/cm$^2$, corresponding to a pulse energy of $0.71\,\mu$J
for a 10$\,\mu$m diameter of the spot size.
Longer Stokes pulses perform similarly well. Complete population 
transfer is achieved by 5$\,$fs and 30$\,$fs pulses with peak
intensities of $3.4 \times 10^{13}$ 
and $3.3 \times 10^{9}\,$W/cm$^2$, corresponding to 
powers of $0.10\,\mu$J and $0.02\,\mu$J, respectively. 
Populating the intermediate state efficiently by the pump pulse is
thus identified to be the limiting step.  

\begin{figure}[tb]
  \centering
  \includegraphics[width=0.99\linewidth]{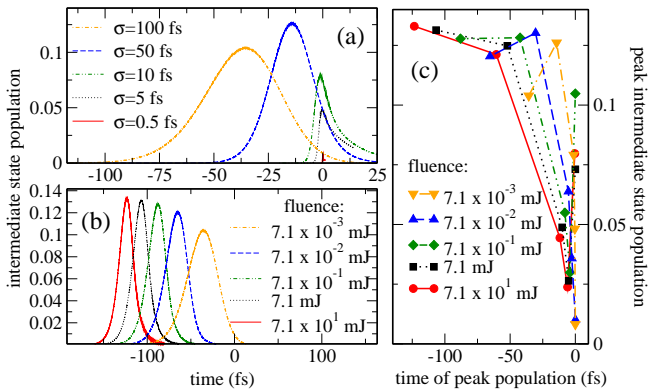}
  \caption{(Color online) 
    Exploring the performance of Gaussian pump pulses:  
    Intermediate state population as a function of time ($t_{0,p}=0$) 
    for (a) different pulse durations  at a pulse energy of 
    $7.1\,\mu$J, and (b) different pulse energies 
    for a FWHM of 100.0$\,$fs. 
    As pulse duration increases, larger intermediate state populations
    can be achieved, although for a given energy the finite lifetime
    acts to drive the intermediate state population down.
    (c) Peak intermediate state population  as a function of 
    the time it is achieved for different pulse
    energies (propagations with TDCIS-HFS-1P).
\label{intlifetimeds}}
\end{figure}
The performance of the pump step is analyzed in
Fig.~\ref{intlifetimeds} for Gaussian pulses with 
different pulse durations (a) and energies (b): 
At a given energy, the maximum intermediate state
population first increases with pulse duration ($\sigma \leq
50\,$fs). This is explained by a better selectivity of 
longer, i.e., spectrally narrower pulses, which avoid populating other  
resonances nearby. However, for very long pulses, the maximum
intermediate state population decreases due to the lifetime of
the intermediate state, which is about 25$\,$fs~\cite{CodlingPR67}. 
%%% lg: addressing comment below
Increasing the pulse energy for a given duration moves the
intermediate state population maximum to earlier times and achieves
larger maxima for a duration of 100$\,$fs.
However, for the smaller powers considered, increasing pulse duration can
lead to a decrease in the population maximum (Fig.~\ref{intlifetimeds}(c)).
The best compromise in terms of pulse power
and performance is found for a 50.0$\,$fs/$71\,\mu$J pulse 
(with peak intensity of $2.4 \times
10^{15}\,$W/cm$^2$) which achieves a maximum intermediate state
population of 0.130, only slightly smaller than the peak at 
0.133 seen in Fig.~\ref{intlifetimeds}(b)
for 100.0$\,$fs/71$\,$mJ ($1.2 \times 10^{18}\,$W/cm$^2$)
pulse. 

\begin{figure}[tb]
\includegraphics[width=0.95\linewidth]{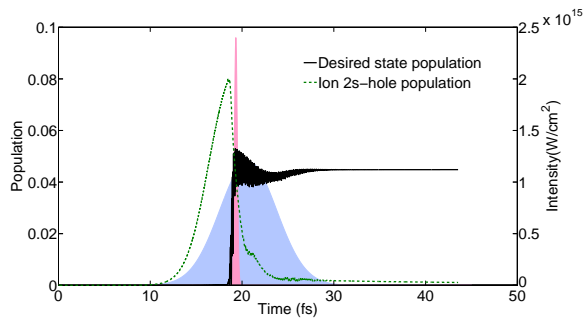}
\caption{(Color online) 2s hole population (dashed green line), 
  corresponding mainly to the intermediate state, 
  and target state populations  (solid black line)
  for a 10$\,$fs/$7.1\,\mu$J pump pulse
  and a 0.5$\,$fs/$0.71\,\mu$J Stokes pulse centered at 18.7$\,$fs
  (peak intensities of 1.2 and $2.4 \times
  10^{15}\,$W/cm$^2$, respectively). 
  Pulse sequences are shown in light blue (pump) and red (Stokes).
\label{propsf10}}
\end{figure}
Improved pulse sequences may now be obtained by setting the Stokes
pulse maximum to coincide with the maximum intermediate
state population, using the pulse energy that balances the  
incidental transfer to undesired states with the intermediate state
lifetime. An example is shown in Fig.~\ref{propsf10}, yielding a Raman
state population of 4.5\%, compared to the 7.9\% maximum intermediate state population.
%%% chr: Please summarize the best results
%%% lg: These are actually different calculations.  In order to determine the
%%% feasability of longer Stokes pulses, I used TDCIS-HFS-1P and the
%%% long pump pulses that do 13% population.  The calculations presented
%%% in Fig. 5 are full TDCIS calculations for shorter pump pulses.
%%% The longer Stokes pulses would probably work the same here, but I
%%% didn't check this.  I'm modifying the following sentences. 
We find final target populations of 13\%, using TDCIS-HFS-1P, when increasing the pump
duration to 50$\,$fs, keeping pump energy and Stokes duration and energy
fixed. Such pump pulse parameters are feasible at free electron laser
(FEL) facilities where it might prove difficult to realize
sub-femtosecond Stokes pulses, however, but longer Stokes pulses may be
employed as discussed above.
When using these sequences of Gaussian pulses as initial guess for
further optimization with Krotov's method,  only small improvements
are found. This suggests these sequences to be already close to
optimal. 

The simplicity of our pump-Stokes sequences raises the question
whether other commonly used  schemes perform equally well. This is not
the case: For example, merely increasing the amplitudes for a set of
simultaneous pulses achieves target populations of 0.2\% at most, for
a pulse intensity of $10^{16}\,$W/cm$^2$, before competing population
transfers to other states and ionization take over and the populations
fall again. This approach also does not provide the phase
control needed to reach arbitrary superposition states. Another
candidate is stimulated Raman adiabatic passage
(STIRAP)~\cite{bergmann1998stirap} where the Stokes pulse precedes the
pump to avoid population of the intermediate state and thus the
detrimental effect of the intermediate state lifetime. However, given
the couplings and energies of coherent xuv Raman excitation in neon, 
adiabatic transfer is found to be possible only when unrealistically long pulses
of hundreds of picoseconds to nanoseconds are employed. 

This work presented a combined coherent quantum control and electronic
structure method which optimizes the population and coherence of an
xuv Raman excitation of an atom. The method is unique in that it
includes the effects of both channel coupling and ionization, together
with a powerful technique of optimal coherent control.  The optimal
control approach yields locally optimal forms of the laser pulses and
also reveals insight into the physical mechanism of the
continuum-mediated coherent Raman excitation. This allows a relatively
simple pathway model to be employed for generation of  sequential
pulse schemes that can be tailored to specific experimental
constraints without loss of fidelity.  

For the Ne example studied here, we found a sequential intermediate
state-Raman state excitation to be optimal and identified the limiting
step to be populating the intermediate resonance from the ground state. 
In contrast, a counterintuitive STIRAP pulse sequence was found to be
unfeasible, due to the small coupling between the ground and 
intermediate state. 
%%% chr: I would shorten the following and replace it by just stating
%%% the best sequence (probably 50fs pump/0.5fs Stokes) and the best
%%% realistic sequendce (probably 50fs pump/5 or 10 fs Stokes)
%Intermediate state populations of 4.8~\% were found to be achievable with 5 fs pulse durations using a peak intensity of $2.4 \times 10^{15}~\textrm{W}/\textrm{cm}^2$.  
%These populations increase with pulse duration to 13~\% at pulse durations of 50 and 100 fs and peak intensities of  $2.4 \times 10^{15} $ and $1.2 \times 10^{16}~\textrm{W}/\textrm{cm}^2$.
% Complete transfer of population from the intermediate to the Raman state could then be achieved with 
%a subsequent Stokes pulse of 0.5 fs FWHM duration and $2.4 \times
%10^{15}~\textrm{W}/\textrm{cm}^2$ peak intensity.
A 50$\,$fs, 7.1$\,\mu$J pump pulse populates the intermediate state up
to 13\%,  and this population can be completely transferred to the
Raman state  with various Stokes pulses, for example with a duration
of 0.5$\,$fs and an energy of 0.71$\,\mu$J, 
%5$\,$fs and 0.10$\,\mu$J, 
or 30$\,$fs and 0.02$\,\mu$J.
%%% lg:  This is really only true at the TDCIS-HFS-1P level, should I
%%% add that a factor of 2 reduction might be expected with full
%%% TDCIS? 
%%% chr: I would add, but in the main text where this result is
%%% mentioned, not in the conclusions
Most important for the realization of coherent multi-dimensional
spectroscopies, we showed that it is possible to excite a
superposition of the Raman and ground states with a controllable
relative phase.  Here the primary control knob was found to be the
carrier envelope phase of the pump pulse. 
The pulse intensities and durations we have determined here are in
principle possible to realize with FELs, in particular
seeded FELs such as FERMI@Elettra that provide the required time resolution to
compete with the resonance lifetime. 
The optimized coherent Raman calculations thus provide design
principles for future FELs, especially in regard to the need for
multiple beamlines in order to implement a multidimensional x-ray
spectroscopy scheme and the choice of frequencies. 

The sequential pump-Stokes scheme illustrated here can readily be
applied to other atoms.   
%BW 0818 I am not sure that the estimates suggested below would work -
%what about the energy differences from other atoms? These would also
%affect the excitation and transfer probabilitiesÉ. 
% LG 0822 I am assuming no detuning, so I think that all that matters
% is the transition dipole to calculate the Rabi frequency ? 
Indeed, order-of-magnitude estimates of required pulse durations
and intensities can be obtained for atoms with isolated resonances,
analogous to the neon 2s-3p resonance utilized here, by comparing the
transition dipole strengths of the desired levels with the neon
levels. More detailed calculations will be required for overlapping
resonances, where longer pulses may be
required to preferentially address one resonance over the other, or
alternate mechanisms for Raman excitation may arise which utilize both
(or many) resonances. 
Such pulses should also work well in molecules, since the intermediate
state is localized (atom-like) and driving the population into this
state  is the limiting step. With this demonstration of the key
elements of state-selective population transfer and excitation of
superposition states, coherent multidimensional spectroscopies capable
of probing the dynamics of valence excitations in molecules via
localized core-hole excitations, such as pump-probe and
coherent x-ray Raman scattering (CXRS)~\cite{tanaka2002coherent}, now
appear feasible.  The present work illustrates the usefulness and
promise of the coherent control approach in bringing techniques such
as CXRS to fruition under the challenging environment of atomic and
molecular ionization continua.

\begin{acknowledgments}
The authors would like to thank Michael Goerz and Daniel Reich for assistance with the Krotov control code, and Stefan Pabst and Robin Santra for help with the XCID TDCIS code.
Additionally, we would like to acknowledge Xuan Li, Bill McCurdy, Dan Haxton, Ali Belkacem and Holger Mueller for helpful discussions.
We acknowledge computational resources obtained under NSF award CHE-1048789 and travel assistance provided by NSF international collaboration grant OISE-1158954 and the DAAD. % for computational resources (Pitzer cluster at UC Berkeley).
\end{acknowledgments}

\bibliography{UltrafastControl}

\end{document}